\documentclass[10pt,%letterpaper,
twocolumn]{article}
\usepackage{ol2}
\usepackage[draft]{hyperref}
\usepackage{amsmath}
\usepackage{amssymb}
\usepackage{graphicx}

\newcommand{\ii}{{\rm i}}
\newcommand{\sinc}{\,{\rm sinc}\,}

\errorstopmode

\begin{document}

\twocolumn[ 

\title{Gouy phase for full-aperture spherical and cylindrical waves}

\author{Tom\'a\v{s} Tyc}

\address{Faculty of Science and Faculty of Informatics, Masaryk University,
  Kotlarska 2, 61137 Brno, Czech Republic }

\begin{abstract}
 We investigate the Gouy phase shift for full-aperture waves converging to a
 focal point from all directions in two and three dimensions. We find a simple
 interpretation for the Gouy phase in this situation and show that it has a
 dramatic effect on reshaping sharply localized pulses.
\end{abstract}


 ]

Gouy phase shift~\cite{Gouy1,Gouy2} has been known for more than a century and
it still attracts a lot of interest in the optical community. Its origin has
been explained in different ways and contexts, for an overview see
Ref.~\cite{Visser10} and the references therein.

Gouy phase is a phase shift of a converging wave obtained for instance by
focusing light with a lens when passing through the focus. It turns out that
the phase change of the wave upon passing through the focal point is smaller by
$\pi$ compared to the situation if a plane wave were propagating instead. In
other words, the local wavelength near the focus is slightly larger than
$\lambda_0=2\pi c/\omega$, the wavelength of a plane wave of the same frequency
$\omega$. One interpretation of this fact is that near the focus the wavevector
has non-negligible transversal components and therefore its longitudinal
component must be somewhat reduced, which increases the local
wavelength~\cite{Feng01}.

Gouy phase is usually discussed for light that propagates more or less in one
direction, as when focusing light with a lens or creating a Gaussian beam.  On
the other hand, recently other geometries of light propagation have been
studied in relation to perfect imaging devices such as Maxwell's fish
eye~\cite{Ulf2009-fisheye,Ma10} or to $4\pi$ microscopy~\cite{Hell94}. There,
light arrives at the focal point from all spatial directions and/or leaves that
point for all directions. In this paper we show that Gouy phase exists in this
situation as well and has a nice and straightforward interpretation. We also
discuss the influence of Gouy phase shift on light pulses passing through the
focal point and show that it has a dramatic effect on the pulse shape in the
two-dimensional case.

Our starting point is the wave equation 
\begin{equation}\label{waveeq}
  c^2\Delta\psi-\psi_{tt}=0
\end{equation}
that can be used for describing different types of waves, e.g. scalar or
electromagnetic waves, in a homogeneous non-dispersive medium. Here $c$ is the
speed of the waves.  Now consider a monochromatic spherical wave of frequency
$\omega$ converging to the focal point at the origin of coordinates. This wave
can be expressed as   
\begin{equation}\label{conv}
   \psi_{\rm in}=a\,\frac{\exp[-\ii kr-\ii\omega t]}r
\end{equation}
with $k=\omega/c$. However, Eq.~(\ref{conv}) is not a solution of
Eq.~(\ref{waveeq}) at the very origin because the left-hand side diverges
there.  Therefore, if there is no absorber (drain) for the radiation at $r=0$,
Eq.~(\ref{conv}) does not represent the full solution of the wave
equation. This has a simple reason: the converging wave cannot just disappear
at the focal point~\cite{Ma10} but it changes there into a diverging wave
$\psi_{\rm out}=b\exp[\ii kr-\ii\omega t]/r$ which must be superimposed
with~(\ref{conv}) to get the full solution. The only way to satisfy the wave
equation at the origin is to set $b=-a$, i.e., add the diverging wave with a
phase shift $\pi$, which gives the total wave
\begin{equation}\label{sinc}
  \psi=\psi_{\rm in}+\psi_{\rm out}=-2\ii ka \sinc(kr)\exp(-\ii\omega t)
\end{equation}
The phase shift of $\pi$ between the two waves at the origin can be interpreted
as the Gouy phase.

\begin{figure}[htb]
\begin{center}
\includegraphics[width=70mm]{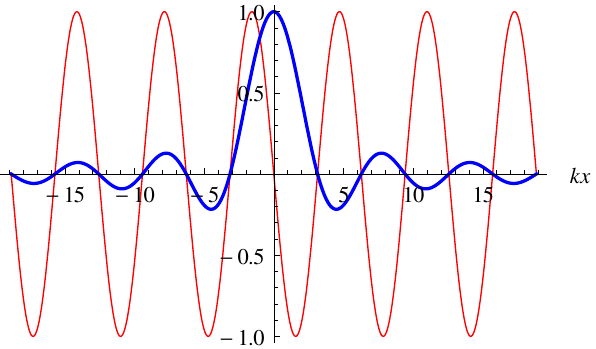}
\end{center}
\caption{Comparison of the function $\sinc kx$ (blue thick curve) and $-\sin kx$
  (red thin curve) describing a superposition of converging and diverging
  spherical waves and a plane wave, respectively, along an axis passing through
  the focal point. The phases match in the region $x<0$ while they differ by
  $\pi$ for $x>0$, which demonstrates the Gouy phase shift.}
\label{sinc-fig}
\end{figure}

To illustrate this in another way, let us have a look more closely at the wave
in Eq.~(\ref{sinc}).  Fig.~\ref{sinc-fig} shows two functions: one is $\sinc
kx$, which represents the spherical wave in Eq.~(\ref{sinc}) (apart
from the global factor $-2\ii ka\exp{(-\ii\omega t)}$) along an $x$-axis
passing through the focal point. The second function is $(-\sin kx)$ and
represents a plane wave of the same frequency propagating along the $x$-axis
whose phase was chosen to coincide with the phase of the sinc wave in the
region $x<0$. Obviously, in the region $x>0$ the phases of the two waves
differ: the sinc wave is delayed by $\pi$ with respect to the plane wave. We
see that a similar thing happens here as near the focus of a lens.  This
demonstrates the Gouy phase for spherical waves in an elegant way, by simply
comparing the graphs of the functions $\sinc kx$ and $\sin kx$. We can also see
that the phase shift occurs on the length scale of the order of the wavelength,
which is consistent with the case of the lens where the Gouy phase shift occurs
on the scale of $\lambda f^2/a^2$ with $f$ denoting the focal distance and $a$
the aperture~\cite{BornWolf}; in our case $f$ and $a$ can be considered to be
of the same order.

\begin{figure}[htb]
\begin{center}
\includegraphics[width=70mm]{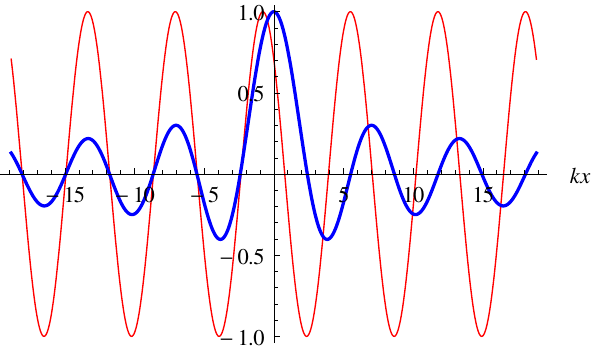}
\end{center}
\caption{Similar as Fig.~\ref{sinc-fig}, but for the 2D situation.  The blue
  thick curve shows the cylindrical wave $J_0(kx)$ and the red thin curve shows
  the plane wave $\sin(kx+3\pi/4)$. The phases match for
  $x\lessapprox-\lambda=-2\pi/k$ but differ by $\pi/2$ for $x\gtrapprox\lambda$
  -- a clear demonstration of the Gouy phase shift in two dimensions.}
\label{bessel-fig}
\end{figure}

A similar consideration can be made for the two-dimensional
rotationally-symmetric waves (or cylindrical waves in 3D). Similarly as before,
a converging wave (created e.g. by the two-dimensional Maxwell's fish
eye~\cite{Ulf2009-fisheye}) turns into a diverging wave at the focal point. The
converging and diverging waves described by Hankel functions are singular at
the origin, but the combined wave satisfies the wave equation at the origin and
is described by the Bessel function of the first kind $J_0(kr)$. Now, similarly
as in the 3D case, we compare this wave with a plane wave of the same
frequency.  Fig.~\ref{bessel-fig} shows the two waves along an $x$-axis passing
through the focal point. The phase of the plane wave was chosen to coincide
with the phase of the cylindrical wave in the region
$x\lessapprox-\lambda=-2\pi/k$ with the help of the asymptotic
formula~\cite{Arfken} for the Bessel function
\begin{equation}
  J_0(kx)\approx \sqrt{\frac2{\pi kx}}\,\cos\left(kx-\frac\pi4\right)\,.
\label{bessel}\end{equation}
Looking at Fig.~\ref{bessel-fig} and/or using Eq.~(\ref{bessel}) again, we see
that in the region $x\gtrapprox\lambda$ the cylindrical wave is delayed by
$\pi/2$ with respect to the plane wave, which is is precisely the Gouy phase
shift in two dimensions.

As the last thing we will demonstrate the effect of Gouy phase on sharply
localized spherical optical pulses converging to a focal point. For convenience
we surround the focal point at the origin by a spherical (in 3D) or circular
(in 2D) mirror of a unit radius. This makes the number of modes countable,
which simplifies expressing the pulses as superpositions of the monochromatic
modes.  We will also set $c=1$ in~Eq.~(\ref{waveeq}), which can always be done
by a suitable choice of units.

Let us start with the 3D case.  For simplicity we will consider scalar waves
now to avoid relatively complicated boundary conditions for the vector
potential~\cite{Sahar11}. Since we are interested in spherically
symmetric pulses only, the spatial parts of the relevant modes depend only on
the radius and are again described by the sinc function as
$\psi_n(r)=\sin(k_nr)/r$, where the numbers $k_n=n\pi$, $n\in\mathbb N$
represent the frequencies of the standing waves that turn to zero at the
mirror. Now consider a pulse emitted at $t=0$ from the origin expressed as the
following superposition of the modes
\begin{equation}
  \psi(r,t) =\frac2r\sum_{n=1}^\infty \sin k_nr\sin k_nt \,.
\label{pulse3d}\end{equation}
With the help of trigonometric identities and Fourier analysis,
Eq.~(\ref{pulse3d}) can be rewritten as
\begin{align}
  \psi(r,t)=& \,\frac1r\sum_{n=0}^\infty\{\cos[k_n(r-t)]-\cos [k_n(r+t)]\}
\label{cos}\\
  =&\,\frac{\Delta(r-t)-\Delta(r+t)}r \,,
\label{delta}
\end{align}
where $\Delta$ denotes the periodic delta-function with period 2,
$\Delta(x)=\sum_{n=-\infty}^\infty\delta(x-2n)$.

We see from Eq.~(\ref{delta}) that in the time interval $t\in(0,1)$ there is a
delta peak propagating from the center to the spherical mirror at $r=1$ that is
reflected from the mirror at time $t=1$, after which it propagates back to the
center with the negative amplitude.  The change of the sign is the result of
the boundary condition at the mirror.  After the converging pulse then reaches
the center at $t=2$, its sign is flipped again and it turns into a positive
diverging pulse. However, this time the sign change is due to Gouy phase. The
position of the peak can also be inferred directly from Eq.~(\ref{cos}): it is
the point where all the first or all the second cosine terms are in phase,
i.e., when either $r-t$ or $r+t$ is an integer.

In two dimensions the situation is more interesting. The rotationally symmetric
modes are the Bessel functions $J_0(k_nr)$ and the frequencies $k_n$ are given
by the boundary condition at the mirror, i.e., at $r=1$, as $J_0(k_{n} )=0$, so
$k_n$ is the $n$-th zero of $J_0$. Consider a pulse
\begin{equation}
  \psi(r,t)=\sum_{n=1}^\infty \sqrt{2\pi k_n}\,J_0(k_nr)
  \sin\left(k_nt+\frac\pi4\right)
\label{pulse-bessel}\end{equation}
Using Eq.~(\ref{bessel}) and the approximate values of $k_n$ derived from it,
\begin{equation}
  k_n \approx \left(n-\frac14\right)\pi\,,
\label{k_n}\end{equation}
we get after some manipulation
\begin{equation}
  \psi(r,t)\approx \frac1{\sqrt r}\sum_{n=1}^\infty 
   \left\{\cos[k_n(r-t)]+\sin[k_n(r+t)]\right\}\,,
\label{deltapulse}\end{equation}
which is valid if $r$ is not too small.  Let us now investigate propagation of
a pulse described by Eq.~(\ref{deltapulse}).  As we will see, the shape of the
pulse is different at different time intervals, therefore we will discuss these
intervals separately.

\begin{figure}
\begin{center}
\includegraphics[width=80mm]{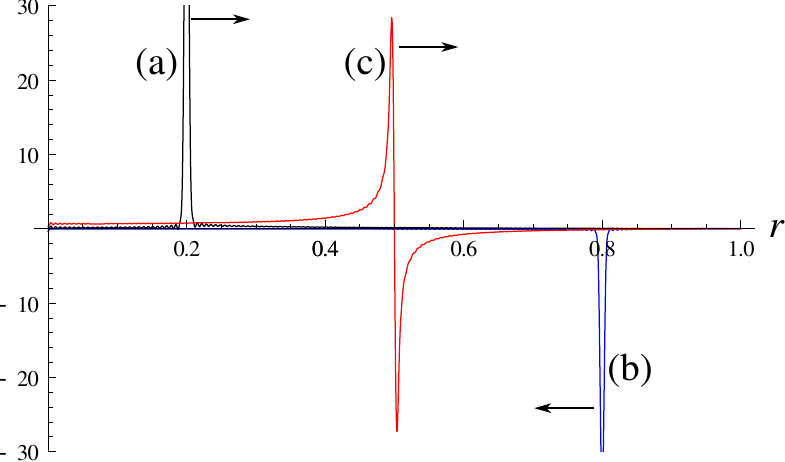}
\end{center}
\caption{Propagation of the pulse described by Eq.~(\ref{pulse-bessel}). The
  sum was truncated at $n=300$ and smoothened by the Gaussian function
  $\exp[-3.5\,(n/300)^2]$ to eliminate rapid oscillations that would occur due
  to the truncation. The pulse is shown at times (a) $t=0.2$ ($\delta$-like
  peak running towards the mirror), (b) $t=1.2$ (negative $\delta$-like peak
  running back to the center), and (c) $t=2.5$ (a cotangent-like peak running
  towards the mirror). Arrows mark the direction of pulse propagation. The
  dramatic change of the peak shape after passing through the center is caused
  by the Gouy phase of $\pi/2$. }
\label{pulse}
\end{figure}

For $0<t<1$ all the cosine terms in Eq.~(\ref{deltapulse}) are in phase at the
point $r=t$. Their superposition therefore forms a peak similar to Dirac
$\delta$-function located at $r=t$ that propagates from the center $r=0$ to the
mirror at $r=1$, see Fig.~\ref{pulse}(a).

To find the behavior of the pulse for $1<t<3$, we make the substitution
$t=2+\tau$ in Eq.~(\ref{deltapulse}) using the approximation~(\ref{k_n}) to the
frequencies $k_n$. This gives
\begin{equation}
  \psi(r,t)\approx \frac1{\sqrt r}\,\sum_{n=1}^\infty
   \left\{-\sin[k_n(r-\tau)]-\cos[k_n(r+\tau)]\right\}\,.
\label{deltapulse2}\end{equation}
For $\tau$ running from $-1$ to 0, which corresponds to $1<t<2$, all cosine
terms in Eq.~(\ref{deltapulse2}) are in phase at $r=-\tau=2-t$. They form a
negative $\delta$-like peak there that runs from the mirror at $r=1$ back to
the origin, see Fig.~\ref{pulse}(b).  The minus sign of this reflected pulse is
again the result of the boundary condition at the mirror.  One might think that
after the pulse passes through the origin at time $t=2$, it will form again a
diverging $\delta$-peak. However, this is incorrect due to the effect of Gouy
phase.  Eq.~(\ref{deltapulse2}) reveals that for $0<\tau<1$ (corresponding to
$2<t<3$) there is no location $r$ at which all cosine terms would have zero
argument. However, now it is the sine terms that can be combined all in phase,
namely at the point $r=\tau$, and their superposition forms a pulse localized
around that point.  But the shape of this pulse is now completely different
from the pulse obtained previously by the cosine terms; instead of a
$\delta$-like peak we now have a double peak resembling the cotangent function,
see Fig.~\ref{pulse}(c), and also the localization of the pulse is much
worse. This dramatic change of the pulse shape is caused by the Gouy phase of
$\pi/2$ that, for each monochromatic component of the pulse, converted a cosine
in Eq.~(\ref{deltapulse}) into a sine in Eq.~(\ref{deltapulse2}). A similar
pulse reshaping has been observed experimentally~\cite{Saari02}.

In conclusion, we have analyzed the Gouy phase for rotationally and spherically
symmetric waves. We have shown that for monochromatic waves the Gouy phase has
a simple meaning that can be visualized by comparing graphs for radially
symmetric solutions of the wave equation with plane wave solutions. We have
also shown that the Gouy phase has a dramatic effect on sharp rotationally
symmetric pulses in 2D whose shape is changed completely after passing the
focal point.

I thank Ulf Leonhardt, Martin Plesch and Michal Lenc for discussions and
acknowledge support of the QUEST programme grant of the Engineering and
Physical Sciences Research Council.

\end{document}